\documentclass{jfm}
\usepackage{amsmath}
\usepackage{amsfonts}
\usepackage{bm}
\usepackage{enumerate}
\usepackage{graphicx}
\usepackage{mathrsfs}
\usepackage{subfigure}
\usepackage{url}
\usepackage{verbatim}
\usepackage{amssymb,graphicx}
\usepackage{hyperref}
\usepackage{verbatim}
\usepackage{xcolor}

\title{Thin gap approximations for microfluidic device design}
\shorttitle{Thin gap approximations for microfluidic device design}

\author{
Lingyun Ding\aff{1}
\corresp{\email{dingly@g.ucla.edu}}
\and Terry Wang\aff{1}
\and Marcus Roper\aff{1,2}
}

\affiliation{
\aff{1}Department of Mathematics, University of California, Los Angeles, CA , USA
\aff{2}Department of Computational Medicine, University of California, Los Angeles, CA , USA
}

\begin{document}
\maketitle

\begin{abstract}
  Over 125 years ago, Henry Selby Hele-Shaw realized that the depth-averaged flow in thin gap geometries can be closely approximated by two-dimensional (2D) potential flow, in a surprising marriage between the theories of viscous-dominated and inviscid flows. Hele-Shaw approximation allows visualization of potential flows over 2D airfoils and also undergirds important discoveries in the dynamics of interfacial instabilities and convection, yet it has found little use in modeling flows in microfluidic devices, although these devices often have thin gap geometries. Here, we derive a Hele-Shaw approximation for the flow in the kinds of thin gap geometries created within microfluidic devices. Using the Method of Weighted Residuals (MWR), we reinterpret the Hele–Shaw approximation as the leading term of an orthogonal polynomial expansion that can be systematically extended to higher-order corrections. The resulting leading-order equation coincides with the previously derived 2D approximations, but our derivation is shorter and more direct. By extending the expansion beyond leading order, we obtain a new reduced model that captures non-parabolic gap-wise velocity profiles and out-of-plane flow effects.  We provide substantial numerical evidence showing that approximate equations can successfully model real microfluidic and inertial-microfluidic device geometries. By reducing three-dimensional (3D) flows to 2D models, our validated model will allow for accelerated device modeling and design.
\end{abstract}
\begin{keywords}
Hele-Shaw approximation; thin-gap flows; microfluidic modeling; inertial effects; 2D approximation; weighted residual methods
\end{keywords}

\section{Introduction }
\begin{figure}
  \centering
      \subfigure[]{
    \includegraphics[width=0.28\linewidth]{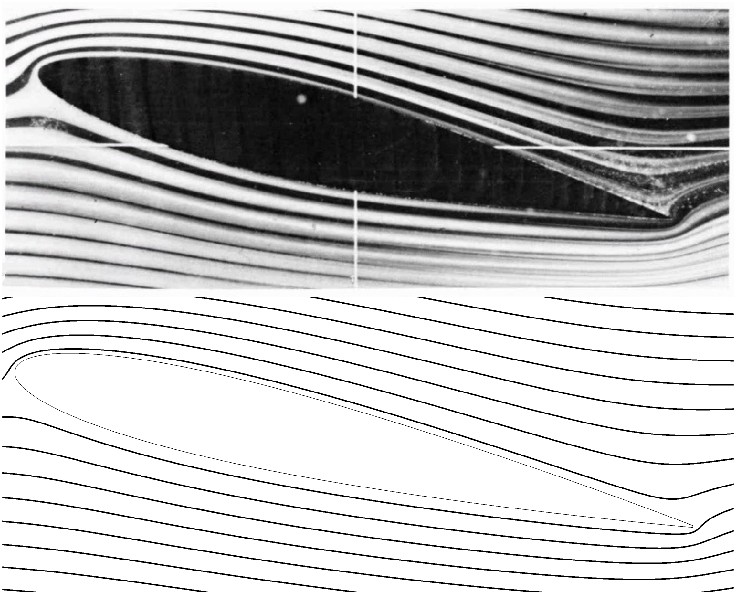}
  }
      \subfigure[]{
    \includegraphics[width=0.33\linewidth]{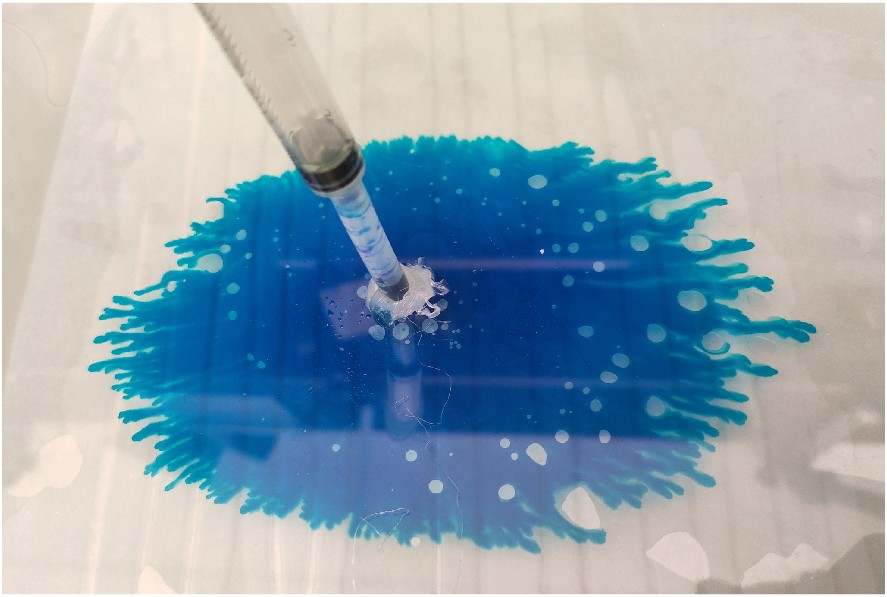}
  }
        \subfigure[]{
    \includegraphics[width=0.32\linewidth]{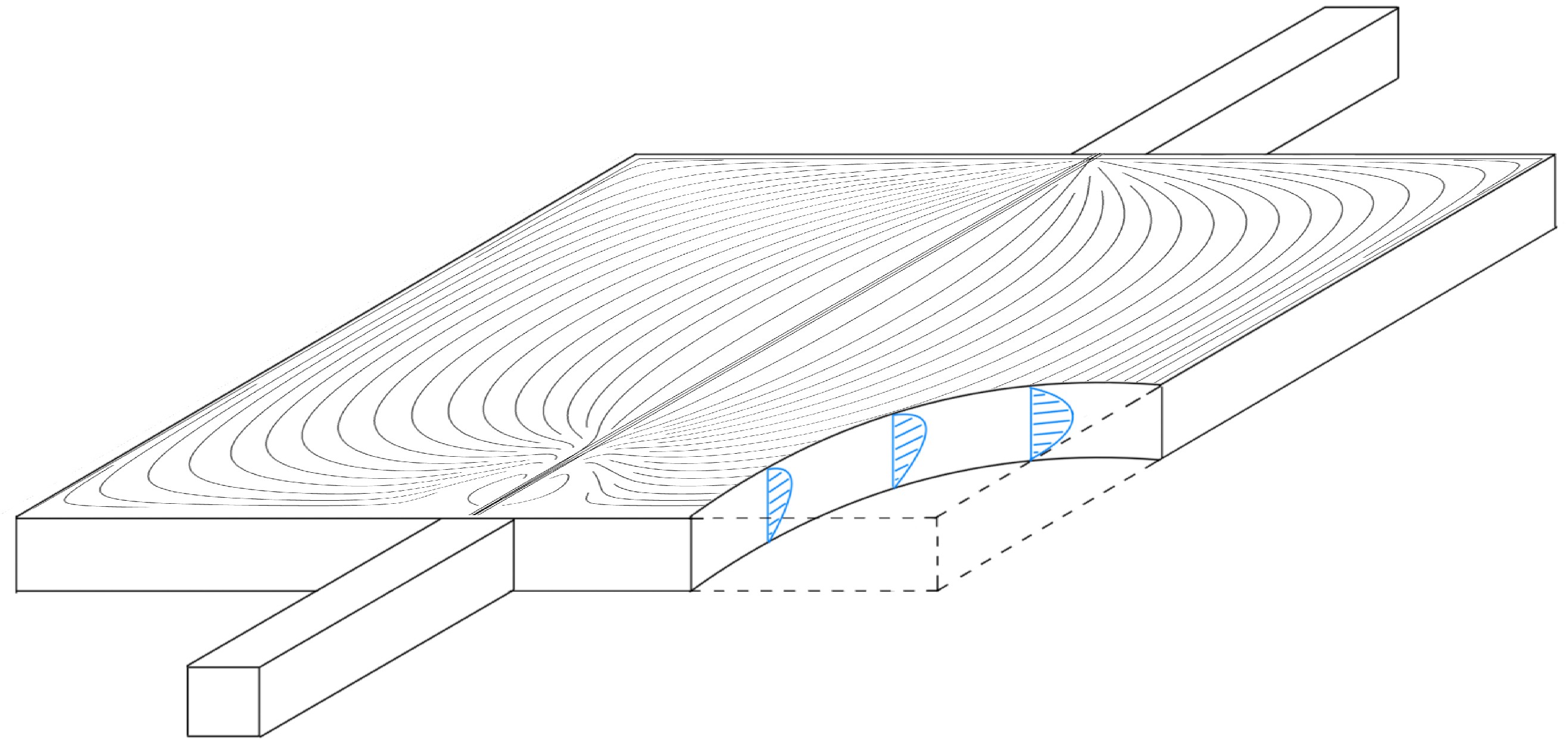}
  }
  \caption[]
  {The Hele–Shaw approximation models thin-gap flows via 2D potential theory. (a) Streamlines around a model airfoil: top—experiment \cite{werle1973hydrodynamic}; bottom—2D potential flow computed in COMSOL 5.4. (b) Thin-gap geometry as a testbed for Saffman–Taylor instability, where a low-viscosity fluid (blue-dyed water) displaces a more viscous one (glycerol). (c) Example microfluidic device with thin gap: a centrifuge-on-a-chip \cite{hur2011high}, where a narrow inlet channel feeds a larger rectangular chamber; the assumed parabolic gapwise velocity profile is shown.
 }
  \label{fig:intro}
\end{figure}

The Hele-Shaw cell, characterized by a narrow gap between two closely spaced parallel plates, is a particularly tractable form of confined flow. Remarkably, flows averaged across the thin gap can be modeled by a 2D potential flow, with the pressure field filling in for the scalar potential. As a result of this tractability, Hele-Shaw geometries can be used to visualize potential flows, such as flow over an airfoil (Fig. \ref{fig:intro}(a,b)), while serving in their own right as useful models for fundamental interfacial instabilities (Fig. \ref{fig:intro}c, \cite{saffman1958penetration,homsy1987viscous}).  
 
 Although Hele-Shaw cells are often thought of as devices for engineering reduced complexity (2D) flows, thin gap domains are also commonly realized in microfluidic devices \cite{stone2004engineering,di2009inertial}, due to the nature of the fabrication methods used to create them. However, the Hele-Shaw approximation can not model no-slip boundary conditions on in-plane boundaries; it has found little use for dissecting the function of existing devices or designing new ones. It remains unanswered whether a different 2D model from Hele-Shaw exists, allowing the key dynamics present within a microfluidic or inertial microfluidic device to be predicted or understood, at far less computational cost than a fully 3D flow model.

The classical derivation of Hele-Shaw's approximation assumes that, within the thin-gap geometry, the flow is planar but varies with $z$, the gap coordinate. Specifically assuming gap boundaries at $z=\pm\frac{h}{2}$, our putative velocity field is parabolic in $z$: $\mathbf{u}(x,y,z) \approx  (u_0(x,y),v_0(x,y),0) \, 6\, (\frac{1}{4}-\frac{z^{2}}{h^2})$, with $u_0$, $v_0$ the depth averaged velocities. Approximating: $\mu \nabla^2 \mathbf{u} \approx \mu \partial^2\mathbf{u}/\partial z^2 = -\frac{12\mu}{h^2} (u_0(x,y),v_0(x,y),0)$, and balancing with the pressure gradient, we arrive at a form of Darcy's equation: $(u_0,v_0) = -\frac{h^2}{12\mu} \nabla p$, which models the pressure-driven flow of fluid through a porous medium, when velocities are averaged over multiple pores
\cite{whitaker1986flow}. Assuming that the flow is incompressible, $\nabla \cdot \mathbf{u} = 0$, we obtain: $-\nabla^2 p = 0$, so that the thin gap flow, $\mathbf{u}$, is a scalar potential flow, and $\frac{h^2 p}{12\mu}$ the harmonic potential. Thus, despite the fact that flow between the plates is viscously dominated, the mean flows are captured by the same potential theory that models high-speed inviscid flows, such as those of ideal fluids over airfoils (Fig. \ref{fig:intro}(a,b)).

The Hele–Shaw approximation captures fluid flows with low complexity, making thin-gap geometries a useful testbed for studying instabilities such as Saffman–Taylor (Fig.~\ref{fig:intro}c) and Rayleigh–Taylor. Thin gaps are also common in microfluidic devices, which manipulate and analyze nanoliter volumes of fluid, but two key deficits limit the approximation’s applicability. It accounts only for the no-through-flow effect of rigid boundaries, treating no-slip as confined to an unmodeled $O(h)$ region near the boundary. It also neglects fluid inertia, which thickens boundary layers to $O(\mathrm{Re} h)$ and creates new regimes that enable sorting and segregation of particles and droplets \cite{di2009inertial}.

Previous work has aimed to incorporate these additional effects into Hele-Shaw-type models. One approach involves adding boundary layer corrections \cite{balsa1998secondary}, though these calculations are often complex and highly geometry-dependent. Another direction is to modify the governing equations themselves. The Darcy-Brinkman equations add in-plane viscous stresses to the Hele-Shaw approximation, \cite{brinkman1949calculation}, and can be used for porous media \cite{tsay1991viscous,howells1974drag}, as well as to thin-gap geometries with empirically determined relative coefficients on the in-plane and in-depth stresses \cite{pepper2010nearby}. There are, by contrast, multiple pathways for incorporating inertial effects. When modeling porous media flow, a nonlinear pressure-flow relation (Darcy-Forchheimer) is commonly imposed \cite{chen2001derivation, guta2010navier}. Conversely, when Darcy's law is arrived at by integrating the flow equations over some thin dimension, inertia-containing terms may likewise be integrated, though there are disagreements between papers on the precise coefficients thereby obtained \cite{desvillettes2008mean,inamdar2020unsteady}.

Here, we arrive at a Hele-Shaw-like approximation via a systematic expansion of the flow field in the gap coordinate. Our approximation naturally includes inertial and in-plane viscous stresses. Comparisons with full 3D simulations show that our approximation accurately models flows in geometries used in real microfluidic devices, including both conventional ($\mathrm{Re}=0$) and inertial ($\mathrm{Re}=1-100$) microfluidic flows.

\section{A new perspective on the Hele-Shaw approximation}
\label{sec:new perspective of the Hele-Shaw approximation}
We consider the Navier–Stokes equations for an incompressible, homogeneous fluid confined within a uniform-thickness gap. The equations are non-dimensionalized using the gap height $h$, so that the vertical coordinate $z$ lies in the interval $-1/2 < z < 1/2$. 
  In the Hele–Shaw limit, the in-plane velocity components have a parabolic profile across the gap. The flow can be approximated by $\mathbf{u}(x,y,z) \approx 3/2\left( 1-4z^2\right) \mathbf{u}_0(x,y)$, where $\mathbf{u}_{0}$ satisfies the no-slip boundary condition on $\mathbf{u}_i$ on any in-plane boundaries. Since $\int\limits_{-\frac{1}{2}}^{\frac{1}{2}}3/2\left( 1-4z^2\right)\mathrm{d}z=1$, $u_0 (x,y)$ is the depth-averaged velocity. Since all velocity components must vanish at the boundaries $z = \pm 1/2$, a natural choice for the higher order vertical basis functions is the family of Gegenbauer (ultraspherical) polynomials with parameter $\alpha = -1/2$, see \cite{belinsky2000integrals}. The Gegenbauer polynomials $C_{n}^{(\alpha)}$ are orthogonal with respect to the weight function $(1-z^{2})^{\alpha - 1/2}$. In particular, $C_{n}^{(1/2)}$ reduces to the Legendre polynomial of degree $n$, and for $n \geq 2$ one has
$C_{n}^{(-\frac{1}{2})} (z)= -\int\limits_{-1}^{z} C_{n-1}^{(\frac{1}{2})} (z) \mathrm{d} z$. The first few polynomials are
\begin{equation}
\begin{aligned}
&C_{2}^{(-\frac{1}{2})} (z)=\frac{1}{2}\left( 1-z^{2} \right), \; C_{3}^{(-\frac{1}{2})}  (z)=\frac{1}{2}z\left( 1-z^{2} \right), \;  C_{4}^{(-\frac{1}{2})}  (z)=\frac{1}{8}\left(5 z^2-1\right)\left( 1-z^{2} \right). 
\end{aligned}
\end{equation}
Since $\lVert C_{n}^{(-\tfrac{1}{2})}(z)\rVert_{\infty} \sim n^{-\tfrac{3}{2}}$, we choose the basis as $a_{n}C_{n+2}^{(-\frac{1}{2})} (2z)$, where $a_{n}$ is a normalization factor chosen so that the basis remains of order one in magnitude. In addition, by the identity $(1-z^{2}) C_n^{(\frac{3}{2})} (z)= (n+1) (n+2)C_{n+2}^{(-\frac{1}{2})} (z)$, it is therefore equivalent to choose $a_{n}(1-4z^{2})C_{n}^{\frac{3}{2}} (2z)$ as the basis, since this differs from our earlier choice only by a multiplicative constant.

Having established an orthogonal family of basis functions that satisfy the no-slip condition at $z=\pm 1/2$, we now represent the three-dimensional velocity field in this basis. Specifically, we expand
\begin{equation}
\begin{aligned}
\mathbf{u} = \sum_{n=0}^\infty \mathbf{u}_n(x, y) a_n C_{n+2}^{(-\frac{1}{2})} (2z),
\end{aligned}
\end{equation}
where $\mathbf{u}_n$ satisfies the no-slip boundary condition on any in-plane boundaries. The subsequent task in constructing the approximation is to specify the corresponding in-plane functions $\mathbf{u}_n(x,y)$ and derive the equations they must satisfy. As a first step, we illustrate the procedure by applying the series expansion to a unidirectional shear flow (rectilinear flow).

\subsection{Poiseuille flow}
\label{sec:Poiseuille flow}
We first consider pressure-driven flow in a channel, aligned with the $x$-axis, with uniform rectangular cross-section $(y,z) \in [-L/2,L/2]\times[-1/2,1/2]$. Supposing an imposed pressure gradient of $G$, and scaling $p_{x}$ by $G$, and velocities by $Gh^2/(12\mu)$, the steady-state solution of Navier-Stokes equations takes the form $\mathbf{u}=(u(y,z),0,0)$, where $u$ satisfies a 2D Poisson equation $(\partial_{y}^{2}+ \partial_{z}^{2}) u = - 12$. In the rectangular channel, this equation admits an exact solution (\cite{cornish1928flow,Boussinesq}). Its series representation is: 
\begin{equation}\label{eq:pressure driven flow exact solution}
\begin{aligned}
u_{exact}=\frac{3}{2} \left( 1-4z^{2} \right)-\frac{48}{\pi^{3}}\sum\limits_{m=0}^{\infty} \frac{(-1)^m}{(1+2m)^{3}}\cos \left( (2m+1)\pi z \right)\frac{\cosh \left( (2m+1)\pi y \right) }{\cosh \left( (2m+1)\pi \frac{L}{2} \right)}.
\end{aligned}
\end{equation}

By substituting the leading order forms of the components of $u\approx \frac{3}{2} (1-4z^{2}) u_{0} (y)$ into the Poisson equation, we obtain the residual:
\begin{equation}\label{eq:pressure driven least square 1}
\begin{aligned}
&r (y,z)=\frac{3}{2} \left( 1-4z^{2} \right) \partial_{y}^{2}u_{0} (y) -12 u_{0} (y) +12 \approx 0,
\end{aligned}
\end{equation}
where the first two terms come from $\partial_{y}^{2}+ \partial_{z}^{2}$, and the third term is the dimensionless pressure gradient. Including additional bases such as $C_{3}^{(-\frac{1}{2})}  (z), C_{4}^{(-\frac{1}{2})}  (z), \ldots$ in our expansion would allow us to eliminate the $z$-dependent terms in Eq.~\eqref{eq:pressure driven least square 1}, but at the expense of introducing a coupled system of equations for the unknown functions $u_i$. Before considering the more complex coupled system, we first seek an optimal closure involving only $u_{0}$, that is, a single equation for $u_{0}$ with constant coefficients. Therefore, we assume the optimal closure is of the form: $a\partial_{y}^{2}u_{0} (y) -12 u_{0} (y) +12=0$ for some constant $a$ to be determined. \footnote{Alternatively, one may assume the more general form  $a \partial_{y}^{2}u_{0} (y) -b u_{0} (y) +c=0$, but, by applying the residual minimization procedure introduced later, one would find that $b = 12$ and $c = 12$, which reduces exactly to the same equation as above.} We can easily solve the equation and obtain the expression of $u_{0} (y)$: 
\begin{equation}
\begin{aligned}
u_{0} (y)= 1-\text{sech}\left(\frac{\sqrt{3} L}{\sqrt{a}}\right) \cosh \left(\frac{2 \sqrt{3} y}{\sqrt{a}}\right).
\end{aligned}
\end{equation}

  To determine $a$, we substitute it into Eq. \eqref{eq:pressure driven least square 1}, obtaining
\begin{equation}\label{eq:pressure driven flow residual}
\begin{aligned}
&r (y,z;a) =u_{0}''(y;a)\left( \frac{3}{2} \left( 1-4z^{2} \right)-a \right).
\end{aligned}
\end{equation}
We choose $a$ to minimize the second factor, integrating the squared residual against a weight function $w(z)$:
\begin{equation} \label{eq:optimalweighted}
  E (a;w) \equiv\int\limits_{-\frac{1}{2}}^{\frac{1}{2}} \left( \frac{3}{2} \left( 1-4z^{2} \right)-a \right)^{2} w(z)\mathrm{d} z.
\end{equation}
Choosing $w(z)=1$, yields $E(a;1)=a^2-2 a+\frac{6}{5}$, which is minimized by $a=1$, thereby recovering the Darcy-Brinkman approximation, and its solution
\begin{equation}\label{eq:pressureDriven weight function 1}
 u\approx \frac{3}{2} (1-4z^{2}) u_{0} (y)=   \frac{3}{2} \left( 1-4z^{2} \right) \left( 1- \cosh \left(2 \sqrt{3} y\right)\text{sech}\left(\sqrt{3} L\right) \right).
\end{equation}
The key step to producing a better approximation is to observe that the space of functions spanned by Gegenbauer polynomials remains the natural space in which to optimize our approximation of Eq. \eqref{eq:pressure driven least square 1}. Since terms in Eq. \eqref{eq:pressure driven least square 1} do not vanish at $z=\pm 1/2$, we multiply the equation by a weight $w (z)= (1-4z^2)$, ensuring the product is contained within that space so that $r(z;a)w (z) = \sum_{n=0}^{\infty} \beta_n C_n^{(-\frac{1}{2})} (2z)$, for some set of coefficients $\beta_m$. Applying Parseval's theorem to this series expansion gives:
\begin{equation}
\begin{aligned}
  &E (a;w) \equiv \int_{-1/2}^{1/2}r(z;a)^2(1-4z^2)\mathrm{d}z= \int_{-1/2}^{1/2}  (1-4z^2)^{-1} \left( r(z;a)(1-4z^2) \right)^{2}\mathrm{d}z\\
  &= \sum_{n=0}^{\infty} \beta_n^2    \int\limits_{-\frac{1}{2}}^{\frac{1}{2}}(1-4z^2)^{-1}  C_n^{(-\frac{1}{2})} (2z)^{2}\mathrm{d} z.
\end{aligned}
\end{equation}
Therefore, our optimal closure consists of minimizing Eq. \eqref{eq:optimalweighted} with $w(z)=(1-4z^2)$. Under this weight, we obtain $E(a;w)=2a^2/3-8 a/5+36/35$, which is minimized when $a=6/5$. This optimal closure leads to a different optimal profile:
\begin{equation}\label{eq:pressureDriven ultraspherical weight function}
\begin{aligned}
 u\approx \frac{3}{2} (1-4z^{2}) u_{0} (y)  = \frac{3}{2}\left(1-4z^{2} \right) \left(1-\cosh(\sqrt{10}y)\operatorname{sech}\left(L\sqrt{\frac{5}{2}}\right)\right)  .
\end{aligned}
\end{equation}
We provide two physical justifications for the parabolic weighting. First, the velocity field vanishes at the boundaries and attains its maximum at the center, so it is natural to assign greater weight near the center and less near the walls. Second, if the solution is independent of $y$, we have 
\begin{equation}
\partial_{z}^{2}u = f, \qquad u|_{z=\pm \tfrac{1}{2}}=0,
\end{equation}
with a general forcing term $f (z)$. The induced volume flux is then
\begin{equation}
F=\int_{-\tfrac{1}{2}}^{\tfrac{1}{2}} u(y)\mathrm{d}z 
= \int_{-\tfrac{1}{2}}^{\tfrac{1}{2}} \tfrac{1}{8} \left( 1-4z^{2} \right) f (z)\mathrm{d}z,
\end{equation}
where the second equality follows from integration by parts.
Thus, when the error between the exact solution and its approximation is measured in terms of the resulting volume flux, the weight $1-4z^{2}$ naturally arises.

\begin{figure}
  \centering
        \subfigure[]{
    \includegraphics[width=0.43\linewidth]{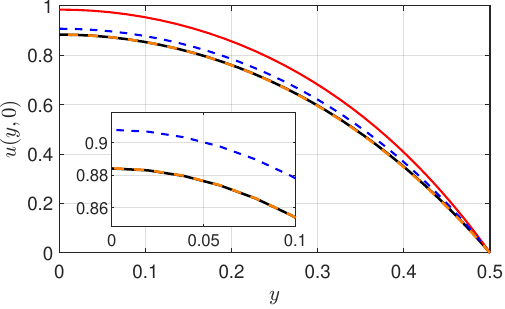}
  }
      \subfigure[]{
    \includegraphics[width=0.43\linewidth]{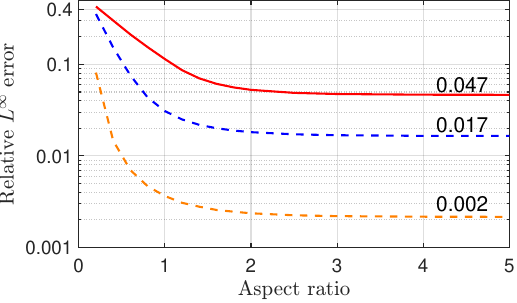}
  }
       \subfigure[]{
    \includegraphics[width=0.43\linewidth]{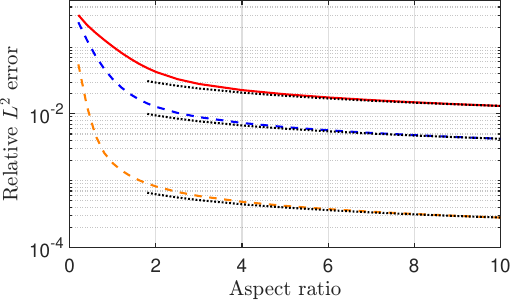}}
     \subfigure[]{
    \includegraphics[width=0.43\linewidth]{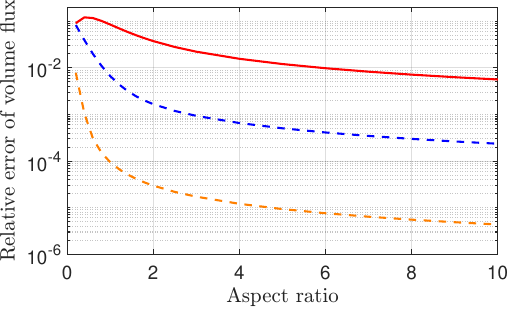}
}
  \caption[]
  {
Validating a 2D approximation for pressure-driven flow in a rectangular channel.
(a) Velocity profile at $z=0$ for $L=1$. Red: Eq. \eqref{eq:pressureDriven weight function 1}; blue dashed: Eq. \eqref{eq:pressureDriven ultraspherical weight function}; orange dashed: higher order approximation from Eq. \eqref{eq:pressureDriven flow 2n order approximation}; black: exact solution. The inset shows a zoomed view of the interval $z \in [0,0.1]$.
(b) Relative error in the $L^\infty$ norm, $\lVert u - u_{\text{app}} \rVert_{\infty} / \lVert u \rVert_{\infty}$, as a function of aspect ratio $L$. Red: Eq. \eqref{eq:pressureDriven weight function 1}; blue dashed: Eq. \eqref{eq:pressureDriven ultraspherical weight function}; orange dashed: Eq. \eqref{eq:pressureDriven flow 2n order approximation}.
(c) Relative error in the $L^2$ norm, $\lVert u - u_{\text{app}} \rVert_{2} / \lVert u \rVert_{2}$, as a function of $L$. Red: Eq. \eqref{eq:pressureDriven weight function 1}; blue dashed: Eq. \eqref{eq:pressureDriven ultraspherical weight function}; orange dashed: Eq. \eqref{eq:pressureDriven flow 2n order approximation}; black dotted: $c L^{1/2}$, illustrating asymptotic scaling.
(d) Relative error in the volume flux as a function of $L$. Red: Eq. \eqref{eq:pressureDriven weight function 1}; blue dashed: Eq. \eqref{eq:pressureDriven ultraspherical weight function}; orange dashed: Eq. \eqref{eq:pressureDriven flow 2n order approximation}.

}
  \label{fig:pipeflow}
\end{figure}

Both approximations derived above are compared against the exact solution. The choice $a=6/5$ consistently yields a more accurate approximation across the entire $y$-domain (Fig. \ref{fig:pipeflow}a) and for all tested values of $L$ (Fig. \ref{fig:pipeflow}b). For example, when $L=1$, the relative errors for $a=1$ and $a=6/5$ are 0.115 and 0.031, respectively. The accuracy of the $a=6/5$ approximation is somewhat surprisingly, given that $L=1$ does not satisfy the classical scale-separation assumption $L \gg 1$. As $L$ increases beyond 1, the relative error decreases for both approximations. 
Though the error measured in the $L^{\infty}$ norm does not vanish in the infinite aspect ratio limit. This is because there always exists a narrow region near the side walls $y = \pm L/2$ where the vertical velocity profile deviates from a parabola due to the no-slip boundary condition. However, as the aspect ratio grows, this region occupies a smaller fraction of the domain, outside of which the pressure-driven flow in the channel approaches that in a two-parallel-plate geometry. When the error is measured in the $L^2$ norm instead, both approximations exhibit an error that decays as $L^{-1/2}$ in the infinite–aspect–ratio limit, as shown in Fig. \ref{fig:pipeflow}(c).

{Another quantity of interest is the volume flux of the approximate flow solution:
\begin{equation}\label{eq:volume flux}
\begin{aligned}
F=\int\limits_{-1}^{1}\int\limits_{-\frac{L}{2}}^{\frac{L}{2}}u_{0} (y)Q_{0} (z)\mathrm{d} y \mathrm{d} z= L-\frac{\sqrt{a} \tanh \left(\frac{\sqrt{3} L}{\sqrt{a}}\right)}{\sqrt{3}}= L-\frac{\sqrt{a}}{\sqrt{3}}+ \mathcal{O} \left( e^{-2\sqrt{\frac{3}{a}}L} \right),
\end{aligned}
\end{equation}
which may be compared to the exact flux formula obtained from \eqref{eq:pressure driven flow exact solution}
\begin{equation}
\begin{aligned}
F=L-\frac{192}{\pi^5}\sum\limits_{m=0}^{\infty} \frac{\tanh \left( \frac{L\pi}{2} (2m+1) \right)}{(2m+1)^5}=L-\frac{186 \zeta (5)}{\pi ^5}+\mathcal{O} \left( e^{-L\pi} \right),
\end{aligned}
\end{equation}
where $\zeta (s)$ is the Riemann zeta function. Figure~\ref{fig:pipeflow}(d) shows the relative error in the volume flux computed using Eq.~\eqref{eq:pressureDriven weight function 1} and Eq.~\eqref{eq:pressureDriven ultraspherical weight function}, demonstrating that the latter achieves consistently better accuracy.  For reference, if side-wall effects are ignored, the solution reduces to
$u(y,z) = \tfrac{3}{2}(1-4z^{2})$, which yields a flux of $F=L$.
In terms of the volume flux, Eq.~\eqref{eq:volume flux} shows that a channel of length $L$ with no-slip side walls is effectively equivalent to a channel of length
$L - \sqrt{2/5}\approx L-0.63246$ with slip side walls. This agrees closely with the asymptotic exact flux, $L-\tfrac{186 \zeta(5)}{\pi ^5}\approx L-0.63025$.

  To probe the optimality of our closure relation more closely, at each aspect ratio, we numerically compute the value of $a$ that minimizes 
\begin{equation}\label{eq:relative error Linfty}
\begin{aligned}
&\frac{\lVert u_{exact} (y,z) -\frac{3}{2}\left( 1-4z^2\right)u_{0}(y;a)\rVert_{\infty} }{\lVert u_{exact} (y,z)\rVert_{\infty}}.
\end{aligned}
\end{equation}
For aspect ratio 1, the value of $a$ is 1.196; for aspect ratio 3, it is 1.188; and for aspect ratio 5, it is 1.184. Hence, the value $a=6/5=1.2$ we obtained by minimizing the residual, closely approximates the optimal error-minimizing values.

We remark that minimizing \eqref{eq:relative error Linfty} requires knowledge of the exact solution to $\Delta u = -12$. For general or more complicated problems, this is impractical. If the exact solution were available, constructing such an approximation would be of limited interest. By contrast, the $L^2$ minimization in \eqref{eq:optimalweighted} does not require solving the governing equation in full, nor even computing $u_{0}(y;a)$. Interestingly, although $u_{0}(y;a)$ depends on $a$, the value of $a$ obtained by minimizing $E(a;w)$ does not necessarily minimize the entire residual $r(y,z;a)$ in the $L^{\infty}$ norm or under other metrics. Nevertheless, it provides a good approximation to the solution. Because this approach yields a reasonably accurate result without requiring the solution of any differential equation, it is computationally attractive.

\subsection{ Method of weighted residuals }

Our closure method follows the general framework of the method of weighted residuals (MWR) \cite{finlayson2013method}, which has been applied to derive reduced-order approximation for thin gap fluid flows ranging from free-surface thin-film flows \cite{kalliadasis2011falling,ruyer2000improved},  Hele–Shaw flows \cite{ruyer2001inertial}, to the thermal convection in a vertical slot \cite{papanicolaou2009galerkin}.

  In the standard application of MWR to obtain a reduced-order model for a differential equation $\mathcal{L}(u) = f$, one first (i) selects a set of basis functions $\left\{ \phi_n(z) \right\}_{n=1}^N$ and approximates the solution as $u(x,z) \approx \sum_{n=1}^N u_n(y) \phi_n(z)$. Next (ii), the residual is then defined as $r(y,z) = \mathcal{L}(u(y,z)) - f(y,z)$. Finally (iii), a set of weight functions $\left\{ w_n(z) \right\}_{n=1}^N$ is chosen, and the residual is projected onto these weights via  $\int_\Omega w_n(z) r(y,z)\mathrm{d} z = 0$, $ j = 1, \dots, N$, yielding a system of $N$ equations for the unknown coefficients ${u_n(y)}$.

We can express both approximations derived in the previous section within the MWR framework. First, we take the weight functions to be $w_{n} = C_{n}^{(1/2)}(2z)$. Owing to the orthogonality of the Gegenbauer polynomials, imposing $\int_\Omega w_n(z) r(y,z)\mathrm{d} z = 0$ is equivalent to expanding the residual in the $C_{n}^{(1/2)}$ basis and setting the corresponding coefficients to zero. Expanding the residual in \eqref{eq:pressure driven least square 1} in terms of $C_{n}^{(1/2)}$ yields
  \begin{equation}
\begin{aligned}
  &r(y,z) = \left( \partial_{y}^{2}u_{0}(y) - 12 u_{0}(y) + 12 \right) C_{0}^{(1/2)}(2z) - \partial_{y}^{2} u_{0}(y) C_{2}^{(1/2)}(2z),\\
  &C_{0}^{(1/2)}(2z)=1, \quad C_{2}^{(1/2)}(2z)=\tfrac{1}{2}(12z^{2}-1).
\end{aligned}
\end{equation}
The coefficient of $C_{0}^{(1/2)}$ recovers the approximation \eqref{eq:pressureDriven weight function 1} with $a = 1$.

Next, we take $w_{n} = (1-4z^{2}) C_{n}^{(3/2)}(2z)$ as the weight functions. Expanding the same residual in the $C_{n}^{(3/2)}$ basis yields
\begin{equation}
\begin{aligned}
  &r(y,z) = \left( \tfrac{6}{5}\partial_{y}^{2} u_{0}(y) - 12 u_{0}(y) + 12 \right) C_{0}^{(3/2)}(2z) - \tfrac{1}{5}\partial_{y}^{2} u_{0}(y) C_{2}^{(3/2)}(2z),\\
 &C_{0}^{(3/2)}(2z)=1, \quad  C_{2}^{(3/2)}(2z)=30z^{2}-\tfrac{3}{2}.
\end{aligned}
\end{equation}
The coefficient of $C_{0}^{(3/2)}$ corresponds exactly to the approximation \eqref{eq:pressureDriven ultraspherical weight function} with $a = 6/5$.

In this sense, the derivation in the previous section is fully consistent with the MWR formulation: the velocity field is expanded in $C_{n}^{(-1/2)}(2z)$, while the residual is projected onto the space spanned by $C_{n}^{(3/2)}(2z)$. The resulting procedure may therefore be interpreted as a Petrov–Galerkin method, since the trial and test spaces are constructed from distinct polynomial families.

Although the optimal closure equation can be obtained within the classical MWR framework by selecting appropriate basis functions and weights, our discussion in the previous section highlights two nontrivial points. First, we demonstrate an alternative to the third step in MWR: instead of integrating the residual against multiple orthogonal weights, we propose a closure equation with unknown parameters, substitute it into the residual, and determine the parameters by minimizing the weighted $L^{2}$ norm. This formulation provides greater flexibility in the closure form and may lead to different relations in other contexts. Second, owing to the completeness of the basis, the systems obtained by projecting the residual onto Legendre polynomials $C_{n}^{(1/2)}$, Gegenbauer polynomials $C_{n}^{(3/2)}$, or even sine functions (as in \cite{pepper2010nearby}), become asymptotically equivalent when many terms are retained in the velocity expansion. However, when only a few terms are used, the choice of weight strongly influences the approximation accuracy. In such cases, one must carefully select both the basis functions and the residual metric to capture the underlying physical behavior of the system.

\subsection{Stokes flow in coaxial flow devices}
\label{sec:Stokes flow in coaxial flow devices}
Under the assumption of Stokes flow, the nonlinear term $\mathbf{u} \cdot \nabla \mathbf{u}$ is again negligible.  The optimal 2D approximation replaces the operator $\Delta$ by $a\Delta_{H}-12$,  where $\Delta_{H}\equiv \partial_{x}^{2}+ \partial_{y}^{2}$, even when the flow is three dimensional. The same choice of $a$ may be justified (residuals resembling Eq. \eqref{eq:optimalweighted} are obtained for each component of the fully 3D Stokes flow). The residual analysis of the continuity equation yields $\partial_{x}u_{0}+\partial_{y}v_{0}=0$.

We test the approximation on a thin-gap but non-rectilinear flow that is commonly realized in microfluidic devices. In \textit{coaxial} flow, three fluid streams, flowing sufficiently slowly that $\mathrm{\mathrm{Re}}=0$, are combined and pass through a narrow aperture (Fig. \ref{fig:coaxialflow}a, \cite{anna2003formation}), emerging as a inner fluid thread sandwiched between the outer fluids. In real coaxial flow devices, the surface tension of the inner fluid thread breaks it up into droplets. We consider a version of the coaxial flow experiment without viscosity contrasts or surface tension, focusing only on the ability of the optimal 2D model to render the geometry of the interface between the inner and outer fluids. We used COMSOL Multiphysics 5.4 to solve both the 3D Stokes equations and the 2D approximate models, and to compare their streamlines with those from the optimal 2D approximation. 
Simulations were performed using the geometry from \cite{anna2003formation}, while varying the mass flux ratio $Q$ between the inner and outer flows.  At a representative value of $Q$, (Fig. \ref{fig:coaxialflow}b), the optimal 2D approximation with $a = 6/5$ closely approximates the interface in both the narrowing and diverging sections of the channel, whereas the classical Hele-Shaw approximation ($a=0$) overpredicts the downstream width of the middle layer, and provides only qualitative agreement in shape. 
\begin{figure}
  \centering
        \subfigure[]{
    \includegraphics[width=0.4\linewidth]{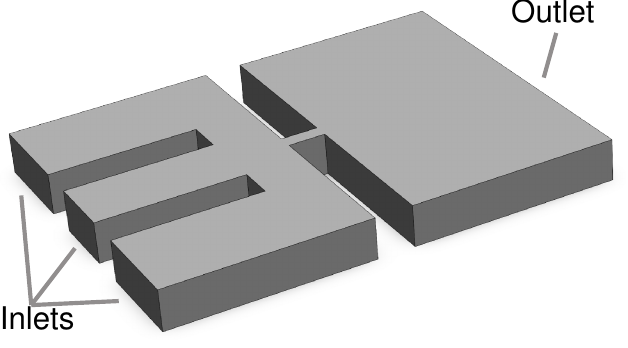}
  }
      \subfigure[]{
    \includegraphics[width=0.4\linewidth]{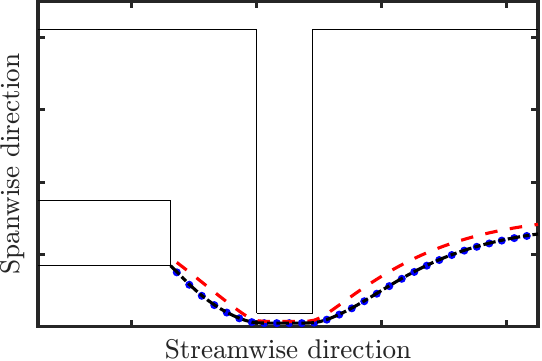}
  }
  \caption[]
  {
 Coaxial flow. (a) Simulation geometry based on \cite{anna2003formation}: left: inlets, right: outlet. The outer inlets have mass flux $Q$, the inner inlet has flux $1$. Inlet and outlet lengths are extended by 20 to ensure fully developed flow.
(b) Dividing streamlines for $Q=1.7$: classical Hele–Shaw (red dashed), derived $a=6/5$ approximation (blue dotted), and 3D simulation (black dashed). }
  \label{fig:coaxialflow}
\end{figure}

The agreement between the 2D approximation and the full 3D computation in the \textit{far} downstream region is expected for two reasons. First, the flow becomes rectilinear, resembling Poiseuille flow—a regime in which our optimal 2D approximation is already known to be accurate (Fig.~\ref{fig:pipeflow}b). Second, mass conservation between the inlets and the outlet constrains the position of the interfaces within the downstream Poiseuille flow. Indeed, it is unsurprising that the classical Hele-Shaw approximation will over-predict the width of the inner-jet (Fig. \ref{fig:coaxialflow}b), because it treats the downstream flow as uniform or plug-like in $y$, omitting the slowdown near the sidewalls of the channel that prevents narrowing of the outer flows. However, the optimal 2D approximation renders the interface accurately even through the narrow orifice separating the inlets and outlet. Although the orifice is relatively narrow (aspect ratio, width: depth $< 0.4$),  and conventional wisdom holds that Hele-Shaw-type models are accurate only when the channel is much wider (aspect ratio $\gtrsim$ 1), the optimal 2D approximation still performs well. Remarkably, the flow remains approximately two-dimensional even within the orifice: the ratio of out-of-plane to in-plane velocity components peaks at just 0.22. This modest degree of three-dimensionality, forced by the thin gap geometry, suggests that the validity of our optimal 2D model extends well beyond the classical limits typically associated with the Hele-Shaw approximation.

\section{Thin gap flows at moderate Reynolds numbers}
\label{sec:Non-zero Reynolds number}
After validating the accuracy of the approximation for Stokes flow (i.e. for thin gap microfluidic devices), we can generalize it to the case of thin gap flows at moderate Reynolds numbers. These conditions are often realized in inertial microfluidic devices, which can be built using the same soft-lithography techniques, and similar dimensions to regular microfluidic devices, but are perfused with high pressure pumps, to create $\sim$ m/s flows (see \cite{di2009inertial}).

For the finite Reynolds number flow, we incorporate an extra term into our closure equation:
\begin{equation}\label{eq:generalNonzeroRe}
\small \mathrm{Re} ( a \partial_{t} \mathbf{u}_{H} + b \mathbf{u}_{H}\cdot \nabla_{H} \mathbf{u}_{H})=  c \Delta_{H} \mathbf{u}_{H}- \frac{d}{H^{2}} \mathbf{u}_{H} - \nabla_{H} p,\; \nabla_{H} \cdot \mathbf{u}_{H}=0, \; \left. \mathbf{u}_{H}\right|_{\partial\Omega}=0.
\end{equation}
Here, $\mathrm{Re} = \tfrac{LU}{\nu}$ is the Reynolds number, with $L$ the characteristic length scale, $U$ the characteristic velocity, and $\nu$ the kinematic viscosity.
The choice of $L$ influences the nondimensionalization. For example, taking $L=\tfrac{h}{2}$ (half the channel depth) yields the vertical domain $z \in [-1,1]$, while taking $L=h$ (the full depth, as in this work) rescales the domain to $z \in [-\tfrac{1}{2}, \tfrac{1}{2}]$. To accommodate both conventions, we denote by $H$ the nondimensionalized gap width.

Following the same weighted residual analysis, using the weight functions $w(z) = 1 -4 z^{2}$ and $w(z) = 1$ yields, respectively:
\begin{subequations}
\begin{align}
&(a,b,c,d)=\left(\frac 65, \frac{54}{35}, \frac 65, 12\right),  \label{eq:partb} \\
&(a,b,c,d)=\left(1, \frac 65, 1, 12\right). \label{eq:parta}
\end{align}
\end{subequations}
Previous work has reported these and similar closure models, arrived at by different approximations and techniques. For example, \cite{plouraboue2002kelvin, yuan2014inertial} used $a = 1$, $b = 1$, $c = 1$, and $d = 12$, while for the case of Stokes flow, \cite{zeng2003brinkman} proposed $a = b = 0$, $c = \frac{12}{\pi^{2}} \approx 1.21$, and $d = 12$. The closure equation with $w(z)=1$ coefficients in Eq \eqref{eq:parta} was derived in \cite{gondret1997shear} by assuming a parabolic flow profile, and averaging the Navier-Stokes equations across the gap.

\cite{ruyer2001inertial} first demonstrated that inertial effects give rise to the coefficient $\tfrac{54}{35}$ in the depth-averaged equation obtained via the MWR. The present work differs in two important respects. First, the coefficient in \cite{ruyer2001inertial} was derived by expanding the velocity field as a sequence of polynomials of the form $z^{j}(1-4z^{2})$ and matching coefficients of like powers of $z$. This non-orthogonal polynomial basis can hinder the derivation of higher-order approximations, leading to more strongly coupled systems with reduced accuracy. Second, the approximation in \cite{ruyer2001inertial} neglects in-plane viscous stresses. Despite this limitation, the model has been widely adopted in studies of fluid instabilities in Hele–Shaw geometries (\cite{chevalier2006inertial}).
\cite{martin2002gravitational} add in-plane viscous stresses with an effective viscosity resulting in identical coefficients to Eq. \eqref{eq:partb}. More recently, \cite{li2021depth} performed an asymptotic analysis, expanding the pressure and velocity fields in powers of $h/L$. By retaining terms up to $O(h^2/L^2)$, discarding terms proportional to the square of the Reynolds number, and several additional approximation assumptions, they obtained the same 2D approximation as \eqref{eq:partb}.  We do not, therefore, make any claim to be the first to arrive at this 2D approximation, but note that optimal closure via MWR provides a direct and physically transparent route to deriving the equation, and its higher order approximations.

\begin{figure}
  \centering
    \includegraphics[width=0.9\linewidth]{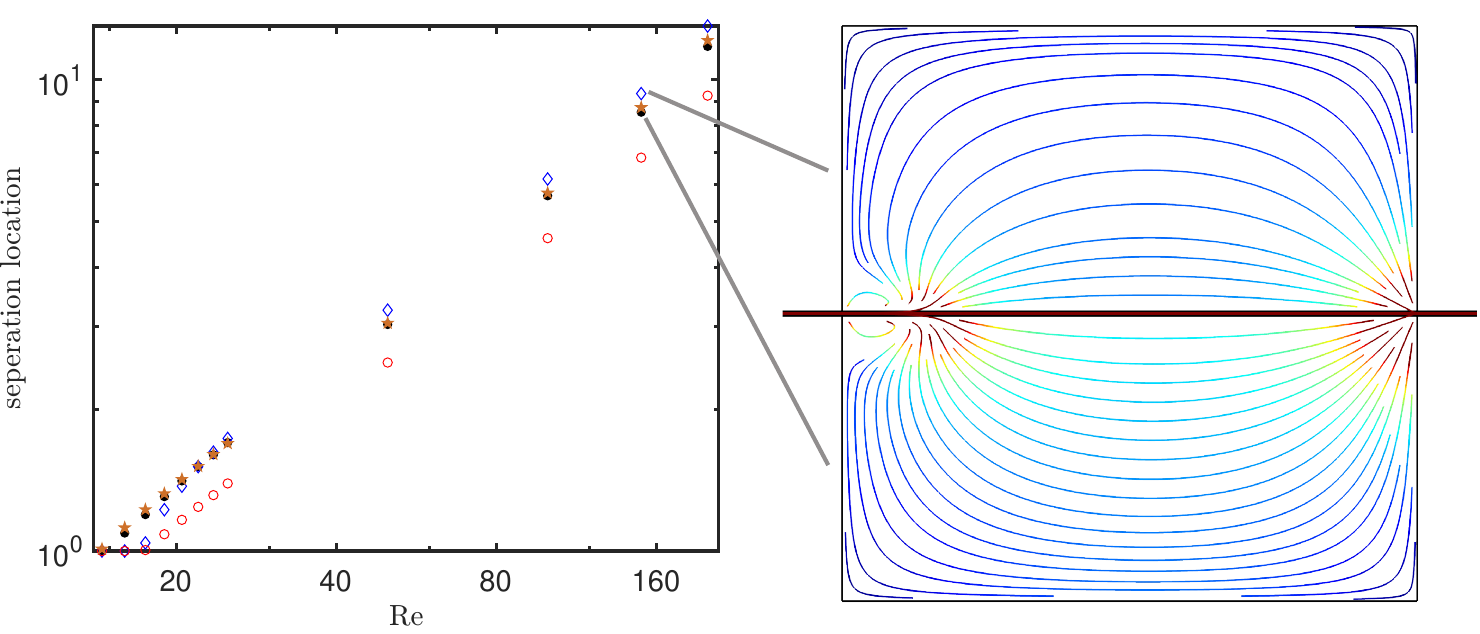}
  \hfill
  \caption[]
  {2D approximation for finite $\mathrm{Re}$ flows in a centrifuge-on-a-chip geometry. Comparison of the separation eddy size in 3D simulation (black), 2D approximation (blue, Eq. \eqref{eq:partb}), second order 2D approximation (orange, Eq. \eqref{eq:2n order flow expansion}), and unweighted residual approximation (red, Eq. \eqref{eq:parta}). Right panels: Streamline patterns colored by velocity magnitude (red = larger, blue = smaller), comparing the 2D approximation (top) with the full 3D simulation (bottom). 

}
  \label{fig:SeparationBubble}
\end{figure}
To compare weighted and unweighted residuals, we compare both 2D approximations with 3D simulations of flow in a centrifuge-on-a-chip (Fig. \ref{fig:intro}c, \cite{hur2011high}), which consists of a rectangular chamber connected to two narrow, elongated channels serving as the inlet and outlet. In our simulations, performed in COMSOL Multiphysics 6.3, the chamber occupies $(x,y,z) \in [-50,50] \times [-50,50] \times [-0.5,0.5]$, while the inlet and outlet have unit square cross-sections. The inlet region of the flow resembles the channel expansion on which \cite{li2021depth} has already validated the 2D approximation \eqref{eq:partb}, but the centrifuge-on-a-chip is a paradigmatic example of the design of inertial microfluidic devices to fractionate particles, such as cells in a blood sample, by size, so centrifuge-on-a-chip flows merit separate attention.

As the Reynolds number $\mathrm{Re} = Uh/\nu$ of the inlet flow is increased, a separation eddy forms at the centrifuge entrance, and this eddy plays a role in trapping of large particles (\cite{paie2017effect}). We measure the eddy size by interpolating the $v=0$ contour to the wall $x=-50$. The 2D approximation \eqref{eq:partb} (left panel, blue diamonds) models the growth of the 3D eddy with $\mathrm{Re}$ (black circles), better than the unweighted residual model (red circles) \eqref{eq:parta}. Though both approximations slightly overpredict the value of $\mathrm{Re}$ at which separation first occurs (17.5 compared to 16 for the fully 3D simulation), the unweighted residual approximation underestimates the separation eddy size by a factor of 16-21\% over the entire $\mathrm{Re}$-range (Fig. \ref{fig:SeparationBubble}), while the error from our 2D approximation \eqref{eq:partb} never exceeds 12\%. Comparing flow fields (Fig.\ref{fig:SeparationBubble}, right panel), we see that the 2D approximation from \eqref{eq:partb} also quantitatively matches the streamlines and velocities obtained in 3D simulations.

\section{Higher order corrections}
\label{sec:Breakdown of the 2D model}
Although the 2D approximation \eqref{eq:partb} shows good accord with 3D simulations, there is mild disagreement in flow features such as eddy size. The framework used to derive the approximation naturally lends itself to determining where and how the approximation breaks down. In forming our approximation, we have adopted the Hele-Shaw approximation of assuming a parabolic velocity profile across the gap, and no fluid velocity in the depth direction. Based on Fig. \ref{fig:SeparationBubble}, we focus particularly on dissecting the features of moderate Reynolds number flows that cause these approximations to break down. We first studied whether the assumption of zero vertical velocity breaks down. Across the entire centrifuge-on-a-chip and for every value of $\mathrm{Re}$ that we simulated, the flow is nearly two-dimensional: the largest value of $|w|_{\infty}/|\mathbf{u}|_{\infty}$ we observe is only 0.0555, obtained in the chamber outlet. 

Next, we examine the validity of the parabolic–velocity assumption. Although a priori estimates for the breakdown of parabolicity can be obtained through linear perturbation analyses of the 3D Navier–Stokes equations in simple geometries, such results remain limited in scope. For example, \cite{plouraboue2002kelvin} analyzed Kelvin–Helmholtz instability in a Hele–Shaw cell and showed that the velocity profile of an inhomogeneous fluid may lose its parabolic character at Reynolds numbers as low as $Re \approx 10$. Because the fluid in our study is homogeneous, the mechanism responsible for this breakdown does not directly apply, and the relevant Reynolds-number threshold may be different.
  
  Our interest is instead in an a posteriori assessment tailored to the Reynolds number and geometry of the microfuge. We quantify deviations from parabolicity by comparing the magnitudes of the first two terms in a Gegenbauer-polynomial expansion of the 3D velocity field.

Specifically,  we interpret the Hele–Shaw approximation as the leading term of an orthogonal polynomial expansion for the velocity. In this, we go beyond reproducing previously derived approximations (\cite{martin2002gravitational,li2021depth}),  generating higher-order corrections on the models for the first time. Including the next order terms, the velocity field and pressure are approximated by
\begin{equation}\label{eq:2n order flow expansion}
\begin{aligned}
&\mathbf{u}_{H}=\frac{3}{2} \left(1-4z^2\right) \left(\mathbf{u}_{0,H}+\left(30 z^2-\frac{3}{2}\right) \mathbf{u}_{2,H}\right),\\
&w=15 z \left(1-4z^2\right)^2w_{2}, \quad p=p_{0}+\left(30z^{2}-\frac{3}{2}\right) p_2,\\
&\mathbf{f}_{H}=\mathbf{f}_{0,H}+\left(30z^{2}-\frac{3}{2}\right) \mathbf{f}_{2,H}, \quad f_{z}=10zf_{2,z},
\end{aligned}
\end{equation}
where $\mathbf{f}_H$ and $f_{z}$ are the body force terms in the horizontal direction and the vertical direction, included here for generality. In the expansion \eqref{eq:2n order flow expansion}, the horizontal velocity component is taken to be even in $z$, which is the most common situation in microfluidic applications.  The expression for $\mathbf{u}_{H}$ follows from retaining both $C_{2}^{(-1/2)}(2z)$ and $C_{4}^{(-1/2)}(2z)$, together with appropriate normalization factors.  Due to the no-slip boundary condition and the continuity equation, both $w$ and $\partial_{z}w$ vanish on the boundary, which determines the gapwise basis function that multiplies $w_{2}$. Moreover, $n\geq4$, $C_n^{(-\frac{3}{2})}(2z)=\frac{-3}{2n-3}\left( C_n^{(-\frac{1}{2})}(2z)-C_{n-2}^{(-\frac{1}{2})}(2z) \right)$ and their derivatives vanish at $z=\pm \frac{1}{2}$, making them a suitable choice of gapwise basis functions for expanding $w$. Accordingly, the vertical momentum equation should be projected onto  $C_n^{(\frac{5}{2})}(2z)$. The coefficient of $p_{2}$ is justified by examining the residual of the leading-order approximation \eqref{eq:pressure driven flow residual}. When $a = 6/5$, the $z$-dependent factor becomes
\begin{equation}
\begin{aligned}
 \frac{3}{2} \left(1-4 z^2\right)-a=-\frac{1}{5} \left(30 z^2-\frac{3}{2}\right)=-\frac{1}{5}C_{2}^{(\frac{3}{2})} (2z).
\end{aligned}
\end{equation}
This choice of basis for $p_{2}$ to efficiently compensate for the dominant residual generated by the other fields. Furthermore, since  $\partial_{z}\left( C_n^{(\frac{3}{2})}(z) \right)=3C_{n-1}^{(\frac{5}{2})}(z)$, the derivative $\partial_{z} p$ naturally produces the $C_{n}^{(5/2)}$ family required to balance the error in the vertical momentum equation. 

Adopting the expansions in \eqref{eq:2n order flow expansion} and  applying the same weighted residual analysis, the continuity equation yields
\begin{equation}
\begin{aligned}
&\nabla_{H} \cdot \mathbf{u}_{0,H}=0, \quad    \nabla_{H} \cdot \mathbf{u}_{2,H}=\frac{20}{3}w_{2}.
\end{aligned}
\end{equation}
The momentum equation in the horizontal direction gives
\begin{equation}\label{eq:2D 2nd order approximation} 
\begin{aligned}
  &\mathrm{Re} \left( \frac{6}{5} \partial_{t} \left(  \mathbf{u}_{0,H} -\frac{3}{7}  \mathbf{u}_{2,H} \right) + \frac{24}{7} \left(-\mathbf{u}_{0,H}+ \frac{54}{11}\mathbf{u}_{2,H}   \right) w_{2} \right.\\
  &\quad\left.   +  \frac{54}{35}\left(  \mathbf{u}_{0,H}-\frac{2}{3}\mathbf{u}_{2,H} \right)\cdot \nabla_{H}  \mathbf{u}_{0,H}+ \frac{36}{35} \left( -\mathbf{u}_{0,H}+\frac{24}{11} \mathbf{u}_{2,H} \right)\cdot \nabla_{H} \mathbf{u}_{2,H}\right)\\
  &=  \frac{6}{5} \Delta_{H} \left( \mathbf{u}_{0,H}-  \frac{3}{7} \mathbf{u}_{2,H} \right)- 12 \mathbf{u}_{0,H} - \nabla_{H} p_{0}+ \mathbf{f}_{0,H},\\
  &\mathrm{Re} \left(\frac{1}{5}\partial_{t}  \left( -\mathbf{u}_{0,H}+4 \mathbf{u}_{2,H} \right)     -\frac{8}{11}\left( \mathbf{u}_{0,H}+ \frac{18}{13}\mathbf{u}_{2,H}   \right) w_{2}   \right.\\
 &\quad\left.   +  \frac{2}{5}\left(  -\mathbf{u}_{0,H}+\frac{24}{11}\mathbf{u}_{2,H} \right)\cdot \nabla_{H}  \mathbf{u}_{0,H}+ \frac{48}{55} \left( \mathbf{u}_{0,H}-\frac{9}{26} \mathbf{u}_{2,H} \right)\cdot \nabla_{H} \mathbf{u}_{2,H}\right)\\
  &=  \frac{1}{5} \Delta_{H} \left( -\mathbf{u}_{0,H}+4\mathbf{u}_{2,H} \right)- 72\mathbf{u}_{2,H} - \nabla_{H} p_{2}+\mathbf{f}_{2,H}, \\
\end{aligned}
\end{equation}
The momentum equation in the vertical direction gives
\begin{equation}
\begin{aligned}
&\mathrm{Re}\partial_{t} w_{2} + \mathrm{Re}\frac{15}{13}\mathbf{u}_{0,H}\cdot   \nabla_{H}  w_{2}  = \Delta_{H} w_{2}- 44w_{2}- \frac{33}{4} p_{2}+\frac{11}{8}f_{2,z}. \\
\end{aligned}
\end{equation}

We first apply the above equation to the unidirectional shear flow problem considered in Section \ref{sec:Poiseuille flow}. We have $u=\frac{3}{2} \left(1-4z^2\right) \left(u_{0}+\left(30 z^2-\frac{3}{2}\right) u_{2}\right)$, where $u_0$ and $u_{2}$ solve
\begin{equation}\label{eq:Poisson equation 2n order approxiamtion}
\begin{aligned}
&0=  \frac{6}{5} \Delta_{H} u_{0}-  \frac{18}{35} \Delta_{H} u_{2}- 12 u_{0} + 12,\\
&0=  -\frac{18}{35} \Delta_{H} u_{0}+ \frac{72}{35} \Delta_{H} u_{2}- \frac{1296}{7} u_{2}. \\
\end{aligned}
\end{equation}
If we set $u_{2}=0$ and only keep the first equation, then the system reduces to the closure equation discussed in Section~\ref{sec:Poiseuille flow}. By an appropriate change of variables, we can reduce equation \eqref{eq:Poisson equation 2n order approxiamtion} to two decoupled equations, leading to the following solution:
\begin{equation}\label{eq:pressureDriven flow 2n order approximation}
\begin{aligned}
&u_{i}(y)= \sum\limits_{j\in \left\{ 0,2 \right\}} a_{0j}a_{ij}\left(  1-\text{sech}\left(\frac{L}{2 \sqrt{\lambda_j}}\right) \cosh \left(\frac{y}{\sqrt{\lambda_j}}\right) \right),\; i=0, 2,\\
&\lambda_{0}=\frac{1}{36} \left(2+\sqrt{\frac{19}{7}}\right)\approx 0.101,\quad \lambda_{2}=\frac{1}{36} \left(2-\sqrt{\frac{19}{7}}\right) \approx 0.00979, \\
&a_{00}=- \sqrt{\frac{1}{2}+\frac{4 \sqrt{\frac{7}{19}}}{5}}\approx -0.992
,\; a_{02}=  \sqrt{\frac{1}{2}-\frac{4 \sqrt{\frac{7}{19}}}{5}}\approx 0.120,\\
&a_{20}=\frac{\sqrt{7}}{6 \sqrt{3}}  \sqrt{\frac{1}{2}-\frac{4 \sqrt{\frac{7}{19}}}{5}}\approx 0.0306,
\;a_{22}=\frac{\sqrt{7}}{6 \sqrt{3}}  \sqrt{\frac{1}{2}+\frac{4 \sqrt{\frac{7}{19}}}{5}} \approx 0.253 .
\end{aligned}
\end{equation}
When $L = 1$, the $L^{\infty}$ error of the second-order approximation is $0.0036$, which is roughly one tenth of the error associated with the leading-order approximation described in Section~\ref{sec:Poiseuille flow}. The second-order approximation, shown as the orange curve in Figure~\ref{fig:pipeflow}(a), is visually indistinguishable from the exact solution. As illustrated in Figure~\ref{fig:pipeflow}(b,c), the errors of the second-order approximation, measured in both the $L^{2}$ and $L^{\infty}$ norms, are substantially smaller than those of the leading-order approximation. Moreover, the $L^{\infty}$ norm of the correction term $\frac{3}{2}(1 - 4 z^{2})(30 z^{2} - \tfrac{3}{2}) u_{2}(y)$ is $0.0249$, which is of the same order of magnitude as the error in the leading-order approximation. This indicates that the correction term provides a practical error estimate of the leading-order approximation.

The flux of this flow approximation \eqref{eq:pressureDriven flow 2n order approximation} is
\begin{equation}\label{eq:pressureDriven flow 2n order approximation flux}
\begin{aligned}
  F&=\int\limits_{-1}^{1}\int\limits_{-\frac{L}{2}}^{\frac{L}{2}}\frac{3}{2} \left(1-4z^2\right) \left(u_{0}+\left(30 z^2-\frac{3}{2}\right) u_{2}\right)\mathrm{d} y \mathrm{d} z\\
  &=\sum\limits_{j\in \left\{ 0,2 \right\}} a_{0j}a_{0j}\left(  L-2 \sqrt{\lambda_{j}}\text{tanh}\left(\frac{L}{2 \sqrt{\lambda_j}}\right)  \right)\\
& = L -\frac{1}{5} \sqrt{\frac{1}{266} \left(9 \sqrt{7}+2618\right)}+ \mathcal{O} \left( e^{-\frac{L}{\sqrt{\lambda_{0}}}} \right).
\end{aligned}
\end{equation}
The constant term in the asymptotic expansion \eqref{eq:pressureDriven flow 2n order approximation flux} obtained from the second order approximation is $0.63029$, matching the exact flux asymptotic expansion to four digits. In contrast, the leading order approximation matches only to two digits. When $L = 1$, the relative error of this flux approximation is
$9.47 \times 10^{-5}$.
For comparison, the relative error of the leading-order flux formula
\eqref{eq:volume flux} is $6.68 \times 10^{-3}$, which is larger
by a factor of $70.6$.
The error of this second-order flux approximation over a range of aspect ratios $L$ is shown as the orange curve in Figure~\ref{fig:pipeflow}(d), demonstrating uniformly higher accuracy across all tested domain sizes.

\begin{figure}
  \centering
         \subfigure[]{
    \includegraphics[width=0.43\linewidth]{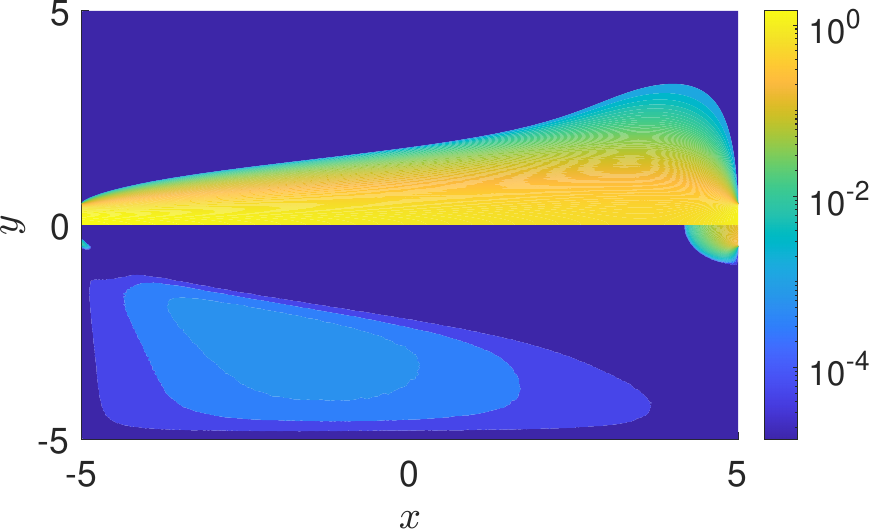}
  }
      \subfigure[]{
    \includegraphics[width=0.43\linewidth]{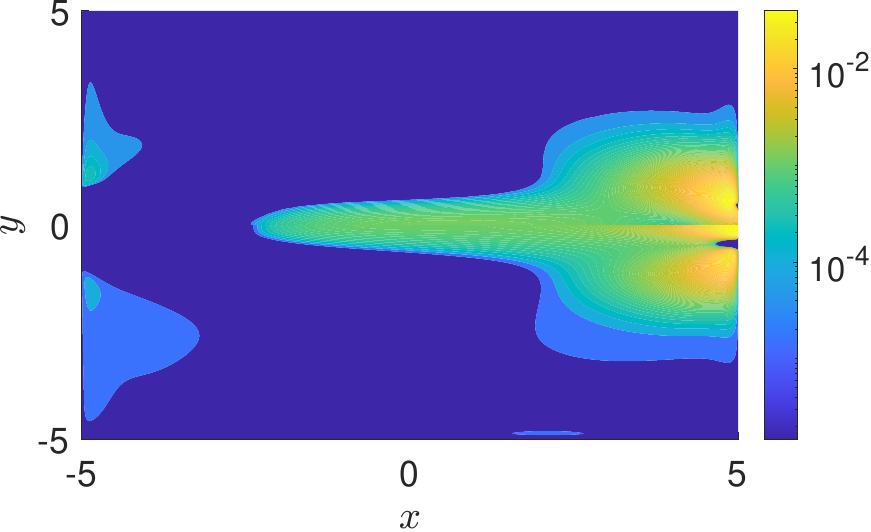}
  }
       \subfigure[]{
    \includegraphics[width=0.42\linewidth]{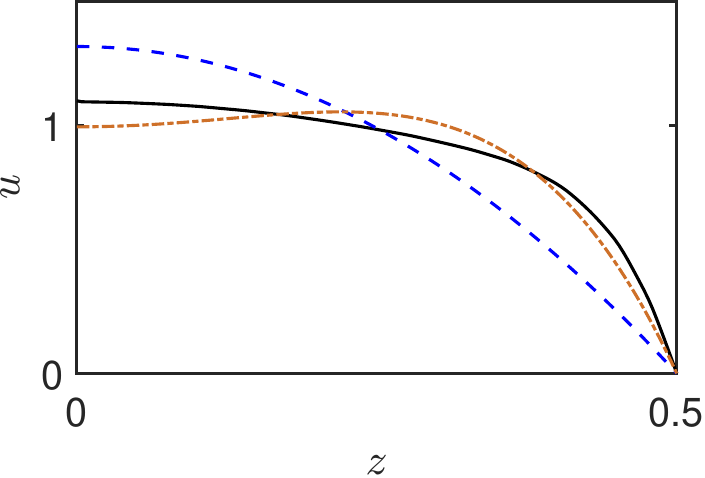}}
     \subfigure[]{
    \includegraphics[width=0.42\linewidth]{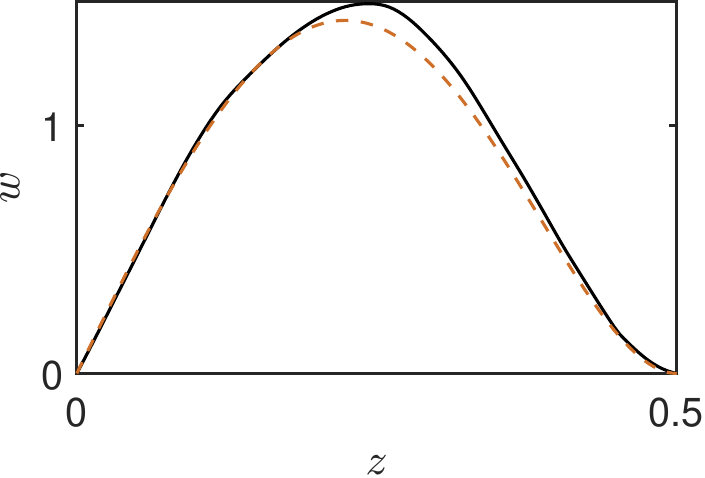}
  }
 \hfill
  \caption[]
  { Simulation of the centrifuge-on-a-chip. The computational domain is $(x,y,z) \in [-5,5] \times [-5,5] \times [-0.5,0.5]$. The simulation uses a Reynolds number of 100 and P2–P1 finite elements. (a) Heatmaps of the $x$-components of $\mathbf u_{0}$ (top) and $\mathbf u_{2}$ (bottom) obtained by projecting the 3D field onto the corresponding basis functions. (b) Heatmaps of $w_{2}$ from the 3D simulation (top) and from the 2D reduced model (bottom). (c) Transects of the $x$-velocity near the outlet ($x = 4.9$, $y = 0.3$): blue dashed curve shows reconstruction using $\mathbf u_{0}$ only; orange dot dashed curve shows reconstruction using $\mathbf u_{0}$ and $\mathbf u_{2}$ via \eqref{eq:2n order flow expansion}; black solid curve shows the full 3D result. (d) Transects of $w$ at $x = -1$, $y = 0.25$ from the 3D simulation (black) and from the 2D reduced model (orange dashed). Vertical velocity units are scaled by $10^{-3}$. }
  \label{fig:heatmap_new}
\end{figure}

Next, we apply the second-order approximation \eqref{eq:2D 2nd order approximation} to the centrifuge-on-a-chip in figure \ref{fig:intro} (c). \eqref{eq:pressureDriven flow 2n order approximation} is used as the boundary condition at the inlet in the simulation of the 2D approximate equation. To ensure the flux is 1 at the inlet, we normalize \eqref{eq:pressureDriven flow 2n order approximation} using  \eqref{eq:pressureDriven flow 2n order approximation flux}.

Second-order corrections are especially valuable when the leading order approximation suffers large errors, or equivalently, when the flow being approximated is highly non-parabolic, or has a cross-gap component. We analyze the centrifuge-on-a-chip to determine where these sources of error occur. In figure \ref{fig:heatmap_new} (a), we plot the magnitude of $\mathbf{u}_{0}$ and $\mathbf{u}_{2}$ calculated by projecting 3D simulation to the basis functions for $\mathbf{u}_{0}$ and $\mathbf{u}_{2}$. The relative size of those components gives a measure of how non-parabolic the flow is. 
In figure \ref{fig:heatmap_new} (b), top panel, we plot the 3D simulated $w$-field.
Taken together, the 3D simulations reveal that the leading-order 2D model becomes inaccurate near the junction between the camber and the outlet. In this region, the gapwise velocity profile departs significantly from the parabolic shape assumed by the leading-order theory, and the cross-gap velocity attains its largest magnitude over the entire domain. These effects indicate the breakdown of the leading-order approximation, and motivate the need for including second-order corrections.

Including the next-order terms alleviates both deficiencies. The bottom panel of figure \ref{fig:heatmap_new}(b), computed from the second-order 2D equations, shows good overall agreement with the full 3D simulation. Figures \ref{fig:heatmap_new}(c) and \ref{fig:heatmap_new}(d) present the in-plane and cross-gap velocity profiles as functions of $z$ at a representative point in the domain, demonstrating close quantitative match between the second-order approximation and the 3D results.

To quantitatively validate the approximation, we compare the produced separation eddy size in figure \ref{fig:SeparationBubble}, finding that the extended 2D approximation capture the eddy size, which is useful for predicting separation performance with almost imperceivable error.

What general benefit can we gained from higher-order approximations?   First, in many microfluidic thin-gap flows, the vertical velocity remains small, making truly 3D flows rare in thin-gap geometries. Conversely, truly 2D flows occur only in special cases, such as Stokes flow in a channel with constant radius of curvature \cite{lauga2004three}. Some applications, however, deliberately induce strong three-dimensionality, e.g. micromixing \cite{stroock2002chaotic}. The higher-order approximation \eqref{eq:2n order flow expansion} provides a nontrivial description of the vertical velocity component, offering valuable insight in these contexts. Second, higher-order terms provide a built-in mechanism for estimating and controlling approximation error. If the estimated error exceeds a chosen tolerance, additional terms can be systematically added to improve accuracy.

While our analysis primarily focused on approximating fluid flows. So long as the domain has a thin dimension, the approach is flexible and broadly applicable. It can be readily extended to other physical quantities, such as solute concentration, electric fields, and magnetic fields, arising in applications like reactive solute transport (\cite{salmon2007transverse}), electrokinetic flows (\cite{lin2008depth,ding2023shear}), and magnetohydrodynamic microfluidics (\cite{qian2009magneto}). The method also allows for generalization of boundary conditions, for example, incorporating no-stress boundaries to model bubble motion (\cite{booth2025motion}) or flows in microchannels with soft walls (\cite{inamdar2020unsteady}).

\section{Acknowledgments}

 The authors acknowledge financial support from NSF award DMS-2009317 (to MR).

 \section{Declaration of interests} The authors report no conflict of interest.

\bibliographystyle{jfm}

\end{document}